\documentclass[aps,prl,showpacs]{revtex4}
\usepackage{amsmath}
\usepackage{graphicx}
\usepackage{bm}
\usepackage{epsfig}

\begin{document}

\title{Scattering in abrupt heterostructures using a position dependent mass
Hamiltonian}
\date{\today}
\author{Ramazan Ko\c{c}}
\email{koc@gantep.edu.tr}
\affiliation{Department of Physics, Faculty of Engineering University of Gaziantep, 27310
Gaziantep, Turkey}
\author{Mehmet Koca}
\email{kocam@squ.edu.om}
\affiliation{Department of Physics, College of Science, Sultan Qaboos University, PO Box
36, Al-Khod 123, Muscat, Sultanate of Oman}
\author{G\"{u}ltekin \c{S}ahino\u{g}lu}
\affiliation{Department of Physics, Faculty of Engineering University of Gaziantep, 27310
Gaziantep, Turkey}

\begin{abstract}
Transmission probabilities of the scattering problem with a position
dependent mass are studied. After sketching the basis of the theory, within
the context of the Schr\"{o}dinger equation for spatially varying effective
mass, the simplest problem, namely, tranmission through a square well
potential with a position dependent mass barrier is studied and its novel
properties are obtained. The solutions presented here may be adventageous in
the design of semiconductor devices.
\end{abstract}

\maketitle

\section{Introduction}

One dimensional quantum wells (QW) and their analysis have played an
increasingly significant role in various applications as well as the
understanding of the properties of a variety of semiconductor devices\cite%
{bastard,tsu,kalotas,weisbuch}. The motivation for studying these problems
is the recent developments in the nanofabrication of semiconductor devices,
where one observes QW with very thin layers\cite{gossard}. The effective
mass of an electron(hole) in the thin layered QW varies with the composition
rate. In such systems, the mass of the electron may change with the
composition rate which depends on the position. Therefore, the corresponding
Schr\"{o}dinger equation should be formulated in a correct form.

Exact and quasi-exact solvability of the position dependent mass (PDM) Schr%
\"{o}dinger equation has been the subject of recent interest\cite%
{koc,alhaidari,milanovic,roy,dekar,koc1,dutra,gonul,koc2}. It provides a
useful model for the description of many physical systems\cite%
{levy,folkues,serra,einevoll,morrow}. Although it has been solved for a
number of potentials and masses, the general solution has not yet been
completed for square well potentials. Here we suggest a model that has been
easily related to the QW structures with various PDM models. We will
demonstrate a number of promising applications of the model.

Potential device applications, as well as purely scientific interest,
provide the motivation for studies of the nature of the transport properties
of the PDM electron through the barriers or wells. For realistic transport
properties in semiconductors, the usual Schr\"{o}dinger equation has to be
replaced by the more general equation\cite{von}:%
\begin{equation}
\left( \frac{1}{4}\left( m^{\alpha }pm^{\beta }pm^{\gamma }+m^{\gamma
}pm^{\beta }pm^{\alpha }\right) +V(z)-E\right) \psi (z)=0  \label{e1}
\end{equation}%
with the constraint over the parameters: $\alpha +\beta +\gamma =-1$. In
applications, the spatial variation of $m$ is either neglected, or,
alternatively various special cases of (\ref{e1}) have been suggested in the
literature \cite{bastard1,gora,duke}. In this article we focus on abrupt
heterostructures. It has been proven \cite{einevoll} that for sharp
heterostructures $\alpha =\gamma $; otherwise the wavefunction is forced to
vanish at the heterojunction boundary which is clearly an unphysical result.

In contrast to the solution of the PDM Schr\"{o}dinger equation including
Coulomb, Morse, harmonic oscillator, etc. type potentials \cite%
{quesne,chen,bagchi,gang}, the study of the PDM Schr\"{o}dinger equation
including a constant potential has not attracted much attention in the
literature. Such quantum systems have been found to be useful in the study
of electronic properties of semiconductors. Generally, analysis of the
scattering problem with PDM is based on the investigation of the simple
problems, and it was pointed out that the transmission probability no longer
tends to unity when incoming energy goes to infinity. The fundamental
question remains open: whether the behavior of the transmission probability
is generic or if it depends on the properties of the mass. To answer this
question one has to obtain a general expression for the transmission
probability by solving (\ref{e1}), for an arbitrary mass.

The rest of the paper is organized as follows. In section 2, we outline a
specific formulation of the exactly solvable PDM Schr\"{o}dinger equation to
derive a general expression for the transmission amplitude of the wave
through the square barrier. In section 3, we apply our model to calculate
the transmission coefficient of the wave through the barrier for various
spatially varying effective masses. Finally, a summary of the work and
conclusions are drawn in section 4.

\section{Theory}

A typical QW structure is composed of a semiconductor thin film embedded
between two semi-infinite semiconductor materials. For a compositional QW,
the well material can be generated by alternate deposition of thin layers.
For example, in a $GaAs/A\ell _{x}Ga_{1-x}As$ QW there exists a wide $GaAs$
well, followed by an $A\ell _{x}Ga_{1-x}As$ barrier and a $GaAs$ narrow
well. The mole fraction $x$ varies along the $z$-axis, therefore the mass of
the electron may vary along the $z$-axis. The simplest model of the QW is
that of a step potential and mass, both showing discontinuities at the same
given point and constant inside and outside the well. Here we suggest a
model by taking into account the spatial variation of the mass inside the
barrier or well. Let us consider a potential barrier of width, $d$. The
structure may be generated by continuously changing the alloy composition $x$
of $A\ell _{x}Ga_{1-x}As$ from $x=0$ to $x=0.32$. The relation between alloy
composition $x$ and coordinate $z$ is given by\cite{zhao1,zhao2}:%
\begin{equation}
x=\frac{0.32z^{2}}{d^{2}}.  \label{ex1}
\end{equation}%
Now we turn our attention to the PDM Schr\"{o}dinger equation (\ref{e1}). As
we mentioned before, the continuity condition forces $\alpha =\gamma =0$ and
$\beta =-1.$ With these choices the PDM Schr\"{o}dinger equation (\ref{e1}),
takes the form:%
\begin{equation}
\left( p\frac{1}{2m}p+V_{0}-E\right) \psi (z)=0,\quad d>z>0  \label{e2}
\end{equation}%
where $V_{0}$ is the constant potential associated with the barrier height,
and $E$ is the energy of the particle. In spite of its simple appearance the
Schr\"{o}dinger equation (\ref{e2}) cannot be solved analytically for
arbitrary $m$. We note here that an exact solution of (\ref{e1}) including a
constant potential can be obtained when $\alpha =\gamma =-1/4$ and $\beta
=-1/2,$ but in this case continuity conditions can not be satisfied. We look
instead at the problem from a different point of view. Instead of the
potential $V_{0}$ let us introduce the following potential \cite{bagchi},
\begin{equation}
V(z)=V_{0}+\frac{\hbar ^{2}}{8m^{2}}\left( m^{\prime \prime }-\frac{%
7m^{\prime 2}}{4m}\right) ,\quad d>z>0  \label{e3}
\end{equation}%
where $m$ is a function of $z$ and $m^{\prime }$ and $m^{\prime \prime }$
denote first and second derivatives of $m$ with respect to $z$. At this
point it is worth mentioning that we will be interested in the potential
which has a less pronounced cusp. Now, the potential resembles a square
barrier or well with smooth walls. The additional term is small compared
with the original potential $V_{0}$ and does not change the shape of the
potential. It is obvious that the conditions are satisfied for smoothly
varying mass. With the potential (\ref{e3}) the Schr\"{o}dinger equation can
be exactly solved with a simple coordinate transformation and the wave
function is given by
\begin{equation}
\psi (z)=\left( C_{1}e^{-ikf(z)}+C_{2}e^{ikf(z)}\right) m^{\frac{1}{4}}
\label{e4}
\end{equation}%
where the function $f(z)$ is defined as $f(z)=\int \sqrt{m}dz$ and $k=\frac{%
\sqrt{2}}{\hbar }(E-V_{0})$.

The results given above can easily be used to solve the Schr\"{o}dinger
equation including well and/or barrier potentials. Let us illustrate our
procedure on a simple example. Consider the potential barrier%
\begin{equation}
V(z)=\left\{
\begin{array}{ll}
0 & 0>z,\quad z>d \\
V_{0}+\frac{\hbar ^{2}}{8m^{2}}\left( m^{\prime \prime }-\frac{7m^{\prime 2}%
}{4m}\right) & d>z>0%
\end{array}%
\right.  \label{e6}
\end{equation}%
with mass barrier%
\begin{equation}
m(z)=\left\{
\begin{array}{ll}
m_{0} & 0>z,\quad z>d \\
m(z) & d>z>0%
\end{array}%
\right.  \label{e7}
\end{equation}%
We assume that the mass of the particle $m_{0}$ is constant outside the
barrier. Mass of the particle inside the barrier $m(z)$ is an arbitrary
function of $z$. The general solution of the Schr\"{o}dinger equation yields:%
\begin{equation}
\psi (z)=\left\{
\begin{array}{ll}
A_{1}e^{ik^{\prime }z}+A_{2}e^{-ik^{\prime }z} & z<0 \\
\left( A_{3}e^{-ikf(z)}+A_{4}e^{ikf(z)}\right) m^{\frac{1}{4}} & d>z>0 \\
A_{5}e^{ik^{\prime }z} & z>d%
\end{array}%
\right.  \label{e8}
\end{equation}%
where $k^{\prime }=\sqrt{2m_{0}E}/\hbar ^{2}$and $A_{i}$ are constants. For
an abrupt heterostructure the continuity conditions are given by\cite%
{einevoll}%
\begin{equation}
m^{\alpha }\psi (z)=\text{continuous},m^{\beta }\frac{d}{dz}m^{\alpha }\psi
(z)=\text{continuous}.  \label{e8a}
\end{equation}%
The transmission coefficient, $T$, and reflection coefficient, $R$, are
defined by%
\begin{equation}
T=\frac{\left| A_{5}\right| ^{2}}{\left| A_{1}\right| ^{2}},\quad R=\frac{%
\left| A_{2}\right| ^{2}}{\left| A_{1}\right| ^{2}},\quad T+R=1  \label{e9}
\end{equation}%
Using elementary quantum mechanical methods, algebraic computation applying
boundary conditions, will lead to the following expression which is related
with the transmission coefficient:%
\begin{eqnarray}
&&\frac{A_{5}}{A_{1}}=  \notag \\
&&\frac{e^{ik^{\prime }d}K_{+}(0)K_{+}^{\ast }(d)e^{ik(f(0)-f(d))}}{%
64kk^{\prime }m_{0}m(0)^{7/4}m(d)^{5/4}f^{\prime }(d)}  \label{e10} \\
&&-\frac{e^{ik^{\prime }d}K_{-}(0)K_{-}^{\ast }(d)e^{-ik(f(0)-f(d))}}{%
64kk^{\prime }m_{0}m(0)^{7/4}m(d)^{5/4}f^{\prime }(d)}  \notag
\end{eqnarray}%
and the coefficient related with the reflection of the wave:%
\begin{equation}
\frac{A_{2}}{A_{1}}=\frac{K_{-}(0)K_{-}^{\ast
}(d)e^{2ikf(d)}-K_{+}(0)K_{+}^{\ast }(d)e^{2ikf(0)}}{K_{+}^{\ast
}(d)K_{-}^{\ast }(d)e^{2ikf(d)}-K_{-}^{\ast }(0)K_{+}^{\ast }(d)e^{2ikf(0)}}
\label{e11}
\end{equation}%
where $K_{\pm }$ are given by
\begin{equation}
K_{\pm }(a)=\left[ 4k^{\prime }m(a)^{2}\pm 4km_{0}m(a)f^{\prime
}(a)-im_{0}m^{\prime }(a)\right]
\end{equation}%
$K_{\pm }^{\ast }(a)$ is conjugate of $K_{\pm }(a).$ The transmission and
reflection coefficients can be computed using the relations (\ref{e8})
through (\ref{e11}). In the following section we will illustrate our model
using some explicit examples.

\section{Examples}

In this section we discuss the dependence of the transmission probability on
the position dependent mass by various choices of the mass $m(z)$. We give
several examples for systems with different position dependent masses. Our
criterion for the selection of masses is that the shape of the original
potential does not change and the square root of $m(z)$ is analytically
integrable. Moreover we made an attempt to include mass functions that are
frequently used in the literature. In order to demonstrate our procedure,
let us begin by considering the following spatially dependent effective
masses found to be useful for studying transport properties in
semiconductors:%
\begin{eqnarray}
m_{a}(z) &=&m_{0}(\sigma +\delta z^{2})  \notag \\
m_{b}(z) &=&m_{0}\sigma e^{\sqrt{\delta }z}  \label{e13} \\
m_{c}(z) &=&m_{0}(\sigma +\tanh (\sqrt{\delta }z))  \notag \\
m_{d}(z) &=&m_{0}\left( \frac{\sqrt{\sigma }+\delta z^{2}}{1+\delta z^{2}}%
\right) ^{2}  \notag
\end{eqnarray}%
where $\delta $ is the length scale parameter and $\sigma $ is a
dimensionless parameter. Through out this section the parameters are chosen $%
\sigma =0.0665,$ $\delta =0.0835,$ and $V_{0}=100meV,$ height of the barrier
and width of the barrier $d=100$ \AA .
\begin{figure*}[tbp]
\begin{center}
\epsfbox{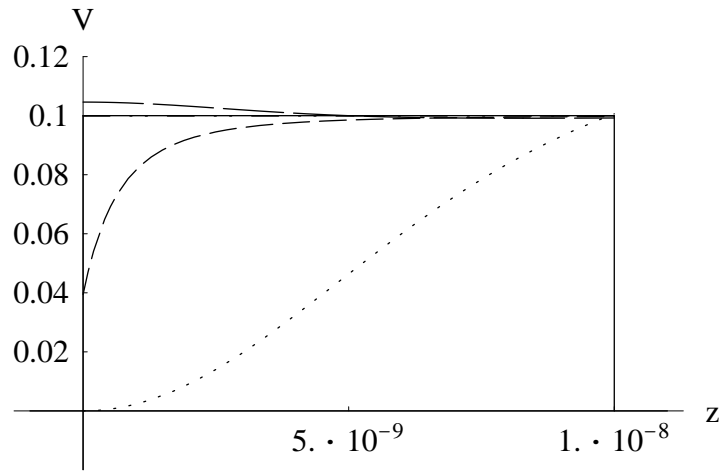}
\end{center}
\caption{Effect of the position dependent mass on the potential profile. The
long dashed line, dashed line and dotted line show the effect of the $%
m_{a},m_{c},$ and $m_{d},$ respectively, on the potential profile. The
change in the potential profile due to $m_{b}(z)$ is plotted with long
dashed lines and it is negligible.}
\label{Figure1}
\end{figure*}

It can be seen from figure \ref{Figure1}, the potential (\ref{e3}) closely
resembles a square barrier with smooth walls for the masses $%
m_{a}(z),m_{b}(z)$ and $m_{c}(z)$. We remark that when the mass rapidly
changes with position $z$, the shape of the potential profile has a
pronounced cusp. The potential (\ref{e3}) which includes the rapidly
changing mass function $m_{d}(z)$ can be plotted as shown in figure \ref%
{Figure1}. We explicitly calculate transmission probability of the
scattering problem employing various physically meaningful spatially varying
effective masses in the following.

\subsection{Mass barrier: $m(z)=\protect\sigma m_{1}$}

\begin{figure*}[tbp]
\begin{center}
\epsfbox{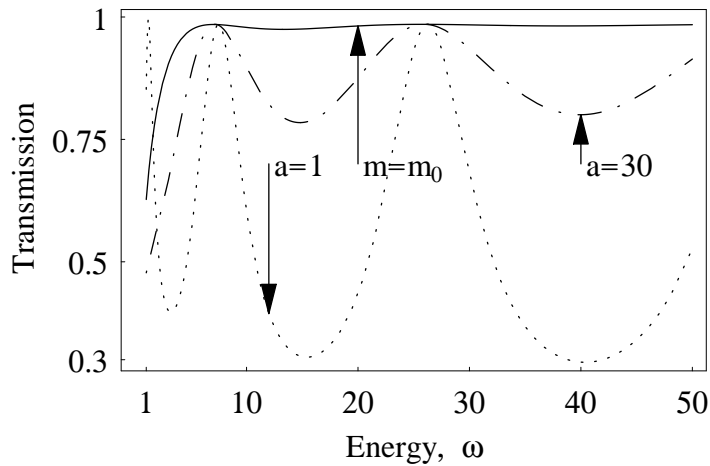}
\end{center}
\caption{The transmission coefficient $T$ for a potential barrier with
height $V_{0}$ and a mass discontinuities. }
\label{Figure2}
\end{figure*}

Consider now a simple mass barrier such that the mass changes at the
potential discontinuities, but inside and outside the barrier it is a
constant. In this case the potential $V_{0}$ remains the same. Since the
tunnelling effect is not qualitatively modified by the mass discontinuity,
we have to leave aside the case where $E<V_{0}.$ In the case $E>V_{0}$ the
calculation for transmission coefficient can easily be done from the
relation (\ref{e10}) and a plot of the transmission probability is
illustrated in figure \ref{Figure2} for various mass ratios. In the plot we
defined the quantities:%
\begin{equation}
\frac{m_{1}}{m_{0}}=a,\quad d=\frac{\pi \hbar }{\sqrt{m_{0}V_{0}}},\quad
\omega =\frac{E}{U_{0}}  \label{e15}
\end{equation}

The graph shows clearly for $m_{0}>m_{1}$ the transmission coefficient no
longer tends to unity when $E$ goes to infinity, but it becomes an
oscillating function of $E$, as is discussed in\cite{levy,sasso}. In figure %
\ref{Figure2}, the curve denoted by $a=1$, corresponds to the plot of
transmission coefficient for $m=0.0665m_{0}.$ This is the conduction band
edge effective mass of the electron in the $GaAs$ structure. In the
following we compute the transmission coefficients for the mass functions
given in (\ref{e13}).

\subsection{Mass Barriers: $m_{a}(z),m_{b}(z),m_{c}(z)$ and $m_{d}(z)$}

\begin{figure*}[tbp]
\begin{center}
\epsfbox{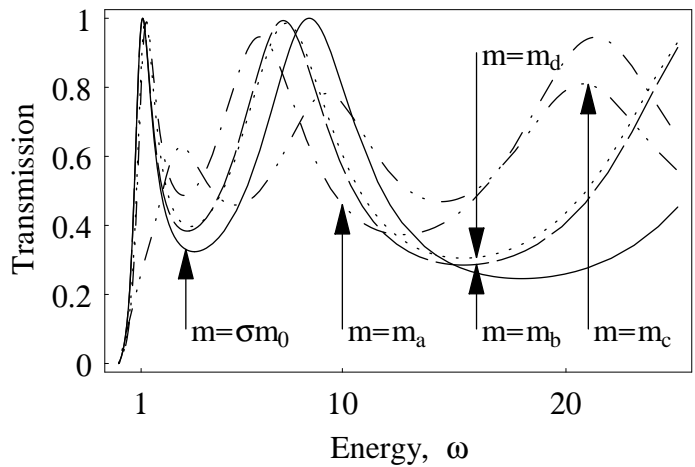}
\end{center}
\par
\caption{The transmission coefficient $T$ for a potential barrier with
height $V_{0}$ and various position dependent mass discontinuities. }
\label{Figure4}
\end{figure*}

The mass functions given in (\ref{e13}) are used in various fields of
physics. We mention here that mass function $m_{a}(z)$, may be useful to
analyze the structures $GaAs/A\ell _{x}Ga_{1-x}As.$ For example the
effective band mass of the electron in the barrier can be written\cite%
{zhao1,qing} as%
\begin{equation}
m(x)=m_{0}\left( 0.0665+0.0835x\right)  \label{e16}
\end{equation}%
The relation between alloy composition $x$ and the coordinate $z$ is given
in (\ref{ex1}). For comparison we calculated transmission coefficients by
using the relations (\ref{e9}) through (\ref{e11}) and they are illustrated
in figure \ref{Figure4}.

\section{Conclusion}

In summary, we have discussed the exact solvability of the PDM Schr\"{o}%
dinger equation including a constant potential. We have recovered a general
expression for the transmission coefficient of the wave through the square
potential barrier. We have presented calculations of transmission
coefficients for various spatially varying effective masses.

Within the framework of the effective mass approximation, in some previous
works\cite{price} the electron was assumed to be confined in a square
infinitely high potential well. In fact, a finite height potential well
model is more realistic for describing the motion of the electron in the QW%
\cite{zheng,lu}. It is obvious that the model described in this article can
easily be modified to study QW structures and superlattices\cite{cruz}.


\begin{thebibliography}{99}
\bibitem{bastard} G. Bastard, \textit{Wave Mechanics Applied to
Semiconductor Heterostructures}, Les Ulis: Editions de Physique, (1988).

\bibitem{tsu} R. Tsu and L. Esaki, Appl. Phys. Lett. 22 (19939 562.

\bibitem{kalotas} T. M. Kalotas and A. R. Lee, Eur. J. Phys. 12 (1991) 275.

\bibitem{weisbuch} C. Weisbuch and B. Vinter, \textit{Quantum Semiconductor
Heterostructures}, New York: Academics, (1993).

\bibitem{gossard} A. C. Gossard, R. C. Miller and W. Weigmann, Surface Sci.
174 (1986) 131.

\bibitem{koc} R. Ko\c{c}, M. Koca and E. K\"{o}rc\"{u}k, J. Phys. A: Math.
Gen. 35 (2002) L527.

\bibitem{alhaidari} A. D. Alhaidari, Int. J. Theor. Phys. 42 (2003) 2999.

\bibitem{milanovic} V. Milanovic and Z. Ikanovic, J.Phys.A:Math. Gen 32
(1999) 7001.

\bibitem{roy} B. Roy and P. Roy J.Phys.A:Math. Gen 35 (2002) 3961.

\bibitem{dekar} L. Dekar, L. Chetouani and T. F. Hammann, J. Math. Phys. 39
(1998) 2551.

\bibitem{koc1} R. Ko\c{c} and H. T\"{u}t\"{u}nc\"{u}ler 2003 Ann.
Phys.(Leipzig) 12 (2003) 684.

\bibitem{dutra} A. de Souza Dutra and C. A. S. Almeida, Phys. Lett. A 275
(2000) 25.

\bibitem{gonul} B\"{u}. G\"{o}n\"{u}l, O. \"{O}zer, Be. G\"{o}n\"{u}l and F.
\"{U}zg\"{u}n, Mod. Phys. Lett. A 17 (2002) 2453.

\bibitem{koc2} R. Ko\c{c} and M. Koca, J. Phys. A: Math. Gen. 36 (2003) 8105.

\bibitem{levy} J. M. Levy-Leblond, Phys. Rev. A\ 52 (1995) 1845; J. M.
Levy-Leblond, Eur. J. Phys. 13 (1992) 215.

\bibitem{folkues} W. M. C. Foulkes and M. Schluter, Phys. Rev. B\ 42 (1990)
11505.

\bibitem{serra} L. I. Serra and E. Lipparini, Europhys. Lett. 40 (1997) 667.

\bibitem{einevoll} G. T. Einevoll, Phys. Rev B 42 (1990) 3497; G. T.
Einevoll and P. C. Hemmer, J. Phys. C: Solid State Phys. 21 (1988) L1193.

\bibitem{morrow} R. A. Morrow, Phys. Rev. B 35 (1987) 8074.

\bibitem{quesne} C. Quesne and V. M. Tkachuk, J. Phys. A: Math. Gen. 37
(2004) 4267.

\bibitem{chen} G. Chen, Chin. Phys. 14 (2005) 460.

\bibitem{bagchi} B. Bagchi, P. Gorain, C. Quesne and R. Roychoudhury, Mod.
Phys. Lett. A 19 (2004) 2765.

\bibitem{gang} C. Gang, Phys. Lett. A 329 (2004) 22.

\bibitem{von} O. von Roos, Phys. Rev. B 27 (1983) 7547.

\bibitem{bastard1} G. Bastard, Phys. Rev. B 24 (1981) 5693.

\bibitem{gora} T. Gora and F. Williams,Phys. Rev. 177 (1969) 1179.

\bibitem{duke} D. J. Ben Daniel and C. B. Duke, Phys. Rev. 152 (1966) 983.

\bibitem{zhao1} F. Q. Zhao, X. X. Liang and S. L. Ban, Eur. Phys. J. B 33
(2003) 3.

\bibitem{zhao2} F. Q. Zhao and X. X. Liang, Chin. Phys. Lett. 19 (2002) 971.

\bibitem{sasso} M. Sassoli de Bianchi and Di. M. Ventra, Solid State Comm.
106 (1998) 249.

\bibitem{qing} Y. Qing and Y. Chu-liang, J. Phys. C: Solid State Phys. 20
(1987) 5125.

\bibitem{price} P. J. Price, Ann. Phys. 133 (1981) 217.

\bibitem{lu} T. Lu and Y. Zheng, Phys. Rev. B 53 (1996) 1438.

\bibitem{zheng} Y. Zheng, T. Lu, Y. Wang, X. Wu, C. Zhang and W. Su,
Semicond. Sci. Thechnol. 12 (1997) 296.

\bibitem{cruz} H. Cruz, A. Hernandez-Cabrera and P. Aceituno, J. Phys. Cond.
Matter 2 (1990) 8953.
\end{thebibliography}
\end{document}